\documentclass[10pt,journal]{IEEEtran}
\usepackage[ansinew]{inputenc}
\usepackage[dvips]{graphicx}
\usepackage[T1]{fontenc}
\usepackage{amsmath,enumerate,amssymb,hhline,verbatim}

\DeclareMathOperator{\Gal}{Gal}

\begin{document}

\title{On the Decay of the Determinants of Multiuser MIMO Lattice Codes}
\author{Jyrki Lahtonen, Roope Vehkalahti, Hsiao-feng (Francis) Lu, Camilla Hollanti, and Emanuele Viterbo}

\maketitle

\newtheorem{definition}{Definition}[section]
\newtheorem{thm}{Theorem}[section]
\newtheorem{proposition}[thm]{Proposition}
\newtheorem{lemma}[thm]{Lemma}
\newtheorem{corollary}[thm]{Corollary}
\newtheorem{exam}{Example}[section]
\newtheorem{conj}{Conjecture}
\newtheorem{remark}{Remark}[section]

\newcommand{\La}{\mathbf{L}}
\newcommand{\h}{{\mathbf h}}
\newcommand{\Z}{{\mathbf Z}}
\newcommand{\R}{{\mathbf R}}
\newcommand{\C}{{\mathbf C}}
\newcommand{\D}{{\mathcal D}}
\newcommand{\F}{{\mathbf F}}
\newcommand{\HH}{{\mathbf H}}
\newcommand{\OO}{{\mathcal O}}
\newcommand{\G}{{\mathcal G}}
\newcommand{\A}{{\mathcal A}}
\newcommand{\B}{{\mathcal B}}
\newcommand{\I}{{\mathcal I}}
\newcommand{\E}{{\mathcal E}}
\newcommand{\PP}{{\mathcal P}}
\newcommand{\Q}{{\mathbf Q}}
\newcommand{\M}{{\mathcal M}}
\newcommand{\separ}{\,\vert\,}
\newcommand{\abs}[1]{\vert #1 \vert}

\begin{abstract}
In a recent work, Coronel \emph{et al.} initiated the study of
the relation between the diversity-multiplexing tradeoff (DMT) performance of a multiuser multiple-input multiple-output (MU-MIMO) lattice code 
and the rate of the decay of the determinants of the code matrix as a function of
the size of the signal constellation. In this note, we state a simple general upper bound
on the decay function and study the promising code proposed by Badr \& Belfiore in close detail.
We derive a lower bound to its decay function based on a classical theorem due to Liouville.
The resulting bound is applicable also to other codes with constructions based on algebraic number theory.
Further, we study an example sequence of small determinants within the Badr--Belfiore code and derive a tighter
upper bound to its decay function. The upper bound has certain conjectural asymptotic uncertainties,
whence we also list the exact bound for several finite data rates.
\end{abstract}

\section{Background and the decay function}

Assume that we are to design a system for $U$ simultaneously transmitting synchronized users, each
transmitting with $n_t$ transmit antennas and, for simplicity so that we end up with square matrices,
over $Un_t$ channel uses. We can describe each user's signals as $n_t\times Un_t$ complex matrices.
A multiuser MIMO signal is then viewed as a $Un_t \times Un_t$ matrix obtained by using the
signals of the individual users as blocks. So each user is occupying $n_t$ rows in this overall transmission matrix.

Any study of DMT questions calls for a scalable set of finite signal constellations.
For the sake of convenience most authors assume that these signal sets of individual users are carved out of
a user specific lattice $\La_j\subset\M_{n_t\times Un_t},j=1,\ldots,U$.

When studying DMT questions it is natural to assume that each user is maximally using the degrees of freedom
available to him/her. Therefore, the lattices of the individual users should be of full rank  $n=2Un_t^2$,
so that each user's signals consist of integral linear combinations of $n$ user specific basis matrices.
A natural scaling parameter is the range of the integer coefficients. We assume that the range is parameterized
by a natural number $N$. Specifically, for the $j$th user, let ${\bf B}_{j,1}, \cdots, {\bf B}_{j,n}$ be a basis for the lattice $\La_j$ of the $j$th user. Then the code associated with the $j$th user is given by
\begin{equation}
{\cal X}_j \ = \ \left\{ X_j \ = \ \sum_{i=1}^n b_i {\bf B}_{n,i} \ : \ b_i \in {\mathbb Z}, -N\le b_i\le N \right\} \label{eq:X_j}
\end{equation}
where each coefficient $b_i,i=1,\ldots,n$, could be freely chosen from the interval $[-N,N]$. Alternatively, $N$-PAM coordinate set could be used. What is essential for our study is that
the set of available signals for a user is of the order $\OO(N^n)$. Our bounds are blind to constant multipliers, so
for example using a spherically shaped signal set instead will not matter. A QAM-oriented reader may then view
encoding as linear dispersion of $\frac{n}{2}=U n_t^2$ independently chosen $N^2$-QAM symbols.
On the other hand, each user may transmit at a different rate or, equivalently, have his/her own rate parameter.
We denote these by $N_1,N_2,\ldots,N_U$, and by $\La_j(N_j)$ the finite signal constellation
obtained by restricting the coefficients of the basis matrices of lattice $\La_j$ to have absolute value
at most $N_j$.

Typical values of $N_j$ are set in terms of the DMT. Assume that the $j$th user transmits at multiplexing gain $r_j$. It in turn means that the size of ${\cal X}_j$ equals
\[
\left| {\cal X}_j \right| \ = \ \text{SNR}^{r_j U n_t}.
\]
Note that by definition $\left| {\cal X}_j \right| \ = \ \left(N_j\right)^n$ and $n=2 U n_t^2$. Hence to achieve multiplexing gain $r_j$ for the $j$th user we have to set
\begin{equation}
N_j \ = \ \text{SNR}^{\frac{r_j}{2 n_t}}. \label{eq:N_j}
\end{equation}
We will say more about the code ${\cal X}_j$ when we examine the DMT performance of the Badr-Belfiore code in Section \ref{sec:4}.

An important class of error events is formed by those, where the receiver is about to make an error in estimating
 every user's signal. This is dominating the system performance in some cases, because with even a relatively
well designed code the channel state may make the received linear combination of individual error vectors
cancel each other out to a significant extent. Such a cancellation is easier to arrange when all the users
are using a large codebook at the same time corresponding to the cases, where all the users are 
transmitting at a relatively high rate. The standard PEP-driven space-time analysis shows that the probability
of such an error event can be related to the determinant of the matrix
$
X:=M(X_1,X_2,\ldots,X_U)=
\left(X_1^T\  X_2^T\ \cdots \ X_U^T\right)^T,
$
where the $n_t\times Un_t$ block $X_j$ from user $\# j$ is a non-zero matrix $\in\La_j$.
The following quantity is then of interest:
$$
D(N_1,N_2,\ldots,N_U)=\min_{X_j\in\La_j(N_j)\setminus\{0\}}\left|\det M(X_1,\ldots,X_U)\right|.
$$
As a natural special case, when all the users are transmitting at exactly the same rate we give special attention
to the function
$$D(N)=D(N_1=N,...,N_U=N).$$
We call both these functions the {\it decay function\/} of the MU-MIMO code $(\La_j),\ j=1,\ldots,U$.
We have tacitly made the assumption that the code designer has provided us with a form of generalized {\it rank criterion} stating
that the matrix $M(X_1,X_2,\ldots,X_U)$ is of a full rank, whenever all the blocks $X_j, j=1,2,\ldots,U$, are non-zero.
Under the rank criterion the decay function will then only take non-zero values.

Is this a misnomer? After all, in the single user MIMO code, lattices within cyclic division algebras
such as the Golden code and the Golden+ code  
enjoy the so called non-vanishing determinant (NVD) property  stating that there is an absolute constant
$\omega>0$ with the property that $D(N)>\omega$ for all values of $N$. In \cite{EKPKL} it was shown
that the NVD-property guarantees the DMT-optimality of a single user code. As we shall see shortly, this is not
possible in the multiuser case, and the determinants will necessarily tend toward zero (under the assumption of the generalized rank criterion\footnote{It has been shown that in order to design DMT-optimal MAC codes, one does not necessarily need to keep up with the generalized rank criterion. It is enough to satisfy the so-called conditional NVD property, see \cite{isit09_LH,isit09_HLV} for details. Naturally for such codes  $D(N)=0$, so they are not of interest here.}).

A natural goal for
the research in MU-MIMO channels would be to have at hand both an explicit criterion guaranteeing the  DMT-optimality of the
family of lattices $(\La_j),j=1,2,\ldots,U$, {\bf and} a class of constructions meeting this criterion.
Some progress in these questions has been made in \cite{Cor,isit09_HLV,isit09_LH}, and it is easy to believe that a condition
expressed in terms of the decay function is also out there \cite{Cor}. In this note we state some interesting results from  \cite{itw10_LVLH} about
the available decay functions of all MU-MIMO lattice codes in general. Based on them, and as the main contribution of this paper, we study in particular the decay function of the code proposed by Badr \& Belfiore (BB-code,  in short).

What kind of decay functions should one expect? We have shown in \cite{itw10_LVLH} that inverse polynomial decay is forced upon us.
All our upper and lower bounds for $D(N)$ are of the form $C N^{-\delta},$ where $\delta>0$ is a real constant.

\begin{definition}
If the decay function of a MU-MIMO code has an upper bound of the form $D(N)\le C_u N^{-\delta_1}$, we say
that the determinants of this code decay with exponent at least $\delta_1$.
Similarly, if the decay function
has a lower bound of the form $D(N)\ge C_\ell N^{-\delta_2}$, then we say that the decay exponent is at most $\delta_2$.
Finally, if for a particular code we find lower and upper bounds of the form
 $
C_\ell N^{-\delta}\le D(N)\le C_u N^{-\delta},
$ 
we say that the determinants of this code \emph{decay with exponent $\delta$}. 
\end{definition}

Of course, asymptotically we prefer a code with
a smaller decay exponent. Equivalently, we say that the code decays with exponent $\delta$ whenever
 $
\lim_{N\rightarrow\infty}-\frac{\log D(N)}{\log N}=\delta.
$

As a word of caution, it is not at all clear that any code has a well defined decay exponent. For example, it may be
that one only gets results for limes superior or limes inferior here.

One of our main results is to show that in the case of the BB-code we can find positive constants $C_\ell$ and
$C_u$ such that
for this promising code we have the bounds
$$
\frac{C_\ell}{N^2}\le D(N) \le \frac{C_u}{N^{5/3}}.
$$
We also give reasoning for our conjecture that, asymptotically, for very large $N$ we expect $\delta=2$. In other
words the determinants decay by the inverse square law. We view this as good news for the BB-code.
Its decay function is under control in this sense. It would not surprise us if further work
on this topic would show that the inverse square decay is essentially the best possible when $U=2$ and $n_t=1$.

Let us now state two theorems from \cite{itw10_LVLH} that will be used for studying the BB-code. Both theorems are based on the pigeon hole principle.

\begin{thm} (Pigeon hole bound, multiantenna case). For any
full-rate $U$-user code, each user transmitting with $n_t$ antennas, there exists a constant
$K>0$ such that
$$
D(N_1=N,N_2=N_3=\cdots=N_U=1)\le \frac{K}{N^{(U-1)n_t}}.
$$
In other words, the determinants of any full-rate $U$ user $n_t$ transmit antenna
code decay with exponent at least $(U-1)n_t$.
\end{thm}

\begin{thm} (Pigeon hole bound, single antenna case)
For any full-rate $U$-user code $(\La_1,\La_2,\ldots,\La_U)$ with $n_t=1$
there exists a constant $K>0$ such that
$$
D(N_1=N,N_2=N_3=\cdots=N_U=1)\le \frac{K}{N^{U-1}}.
$$
In other words the determinants of any single transmit antenna full-rate $U$-user code decay with exponent $\delta\ge U-1$.
\end{thm}

\section{A lower bound to the decay of the Badr--Belfiore code}

Let us recall the code construction from \cite{BB2}. See also
 equation (49) in \cite{Cor}.
This promising code is expressed in terms of certain algebraic number fields.
Everything happens inside the field  $E=\Q(i,\sqrt5)$. We shall also encounter
its subfields $F_1=\Q(i)$, $F_2=\Q(\sqrt 5)$ and $F_3=\Q(i\sqrt5)$.
The respective rings of algebraic integers of these quadratic fields are
$\OO_1=\Z[i]$, $\OO_2=\Z[\tau]$ and $\OO_3=\Z[i\sqrt5],$ where $\tau=(1+\sqrt5)/2$
is the golden ratio. The ring of integers of $\OO_E=\Z[i,\tau]$ then consists
of numbers of the form $(a+bi)+(c+di)\tau$, where $a,b,c,d$ are any rational integers.

The Galois group $G=\Gal(E/\Q)$ has four elements: $1_G; \rho: i\mapsto -i,\ \sqrt5 \mapsto\sqrt5$;\ $\sigma: i\mapsto i,\ \sqrt5 \mapsto -\sqrt5$, and $\mu=\sigma\rho=\rho\sigma:\ i\mapsto -i,\ \sqrt5 \mapsto -\sqrt5$. The respective fixed fields of $\sigma,\rho$ and
$\mu$ are $F_1$, $F_2$, and $F_3$.

We are now ready to describe the BB-code. It fits into our general
framework with parameters $U=2$ and $n_t=1$ so it is a single-antenna two-user code.
Both users linearly disperse two Gaussian integers. User $\#j$ first combines the Gaussian
integers $z_{1j},z_{2j}\in\OO_1$
into an element $x_j=z_{1j}+z_{2j}\tau$ of the ring $\OO_E$. Then their methods differ a little bit,
and user $\#1$ transmits the vector $(x_1,\sigma(x_1))$, whereas the user $\#2$ transmits the vector
$(\gamma x_2,\sigma(x_2))$. In \cite{BB2} and \cite{Cor} it is explained that the choice $\gamma=i$
results in a code that satisfies the generalized rank criterion. In other words, the composite matrix
\begin{equation}
X=
\left(
\begin{array}{cc}
x_1 & \sigma(x_1)\\
\gamma x_2 & \sigma(x_2)
\end{array}\right)\label{eq:Xbb}
\end{equation}
is invertible, whenever both $x_1$ and $x_2$ are non-zero.

Following \cite{Cor} we shall study the determinant
$$
\det(X)=x-\gamma\sigma(x)=x-i\sigma(x),
$$
where $x=x_1\sigma(x_2)$.
Next we describe our finite constellations more precisely.
Let $(a_j,b_j,c_j,d_j)\in\Z^4$ correspond to the signal transmitted by user $\#j$, $j=1$ or $j=2$.
In other words, $z_{1j}=a_j+i b_j$ and $z_{2j}=c_j+id_j$. Then the constellations $\La_1(N)$ and
$\La_2(N)$ are obtained from the above constructions by restricting the integer coefficients
$a_j,b_j,c_j,d_j$ ($j=1$ or $2$) into the range $[-N,N]$.

The determinant will now be of the form
$$
\det (X)= (R + S\tau) + (T + V\tau) i,
$$
where $R,S,T,V$ are quadratic homogeneous polynomials with integer coefficients in the 8 integer
unknowns $a_1,a_2,b_1,b_2,c_1,c_2,d_1,d_2$. The result stating that $\det (X)\neq0$ can be rewritten
in the form that these four polynomials cannot vanish simultaneously, unless the input from one of the users
is all zeros.
As we shall see, our estimate on the decay rate will depend
on the size of the integer coefficients $R,S,T,V$. We note the following obvious lemma without proof.

\begin{lemma}\label{coefficientsize}
There exists a constant $K_1>0$ such that for all $x_1\in\La_1(N_1)$, $x_2\in \La_2(N_2)$ we have
the upper bounds 
$
|S|<K_1 N_1N_2$ and $|V|<K_1 N_1N_2.
$
\end{lemma}

We remark here that further limiting the choices of the inputs of individual users (for example to
the ideal of the ring of integers of $E$ used in the construction of the Golden code) amounts
to placing a family of congruences that the input vector $(a,b,c,d)$ must satisfy. This will not change anything
in what follows. After all, then the desired single user constellation will be a subset of
a set of the form $\La(\alpha N)$, where $\alpha>$ is a constant that does not depend on $N$. Thus our estimates
will also be valid for such constellations, because the contribution  from $\alpha$ can be absorbed into
the coefficient $K_1$ (by replacing it with another positive constant). Neither will replacing $i$ with another non-norm element $\gamma$ affect our conclusions --- albeit naturally all the calculations have to be carried out separately for each $\gamma$.

We already know from the pigeon hole bound that for some constant $C$ the decay function of the
BB-code has an upper bound of the form $D(N_1,N_2)\le K/\max\{N_1,N_2\}$
for some constant $K>0$, i.e., the determinant decays with exponent at least $\delta\ge1$.

For a badly chosen code the decay could be very fast, indeed. We shall next show that the number theoretic
structure of the BB-code can be used to derive an inverse polynomial lower bound too. Thus
this code is promising in the sense that it belongs to a class of MU-MIMO codes with inverse polynomial decay.

\subsection {Approximating $\tau$ by rational numbers}

It is known that it is impossible to approximate algebraic integers too well by rational
numbers in the sense made precise by the following result by Liouville\footnote{For more general real algebraic numbers one should asymptotically use a deep result due to
K. F. Roth stating that the exponent $n$ can be replaced with an exponent of the form $2+\epsilon$
for any $\epsilon>0$. The price one pays when doing this is that one no longer has any means of
estimating the constant in the numerator. For asymptotic work Roth's result is obviously superior.
}. 
Similar methods have been used  in e.g. \cite{DB, DTB}.

\begin{thm} (\cite[p.146]{Apo}, Liouville's approximation theorem)
Let $\theta$ be a real algebraic number of degree $n\ge2$. Then there is a positive
constant $C(\theta)$, depending only on $\theta$, such that for all integers $h$
and $k$ with $k>0$ we have
$$
\left\vert\theta-\frac{h}{k}\right\vert>\frac{C(\theta)}{k^n}.
$$
\end{thm}

As an immediate corollary we get the following.

\begin{corollary}\label{tauapprox}
There exists a constant $C$ such that for all integers $h$ and $k$ with $k>0$ we have
 $
|k\tau -h|>\frac{C}{k}.
$
\end{corollary}

As a corollary of Lemma \ref{coefficientsize} and Corollary \ref{tauapprox}  we get our main result:

\begin{thm}\label{mainresult} There exists a constant $K>0$ such that for all sufficiently large $N_1,N_2$ we have
 $
D(N_1,N_2)\ge\frac{K}{N_1N_2}.
$ 
In particular as $N\rightarrow\infty$, we get a decay estimate
 $
D(N)\ge\frac{K}{N^2}.
$ 
In other words the decay of the BB-code has a lower bound corresponding to
an estimate of the decay exponent $\delta\le 2$.
\end{thm}

We want to remark that the result of Corollary \ref{tauapprox} is essentially
the best possible. For example, it is impossible to replace the exponent $1$ of the parameter $k$
in the denominator with a larger number. This is because a simple application of the pigeon hole
principle tell us that there are infinitely many integer pairs $(h,k)$ such that
 $
|k\theta -h|<\frac{1}{k}
$ 
for {\em any} irrational real number $\theta$.

\section{More on the decay exponent of the Badr--Belfiore code}

We already know that the decay exponent of the BB-code is in the interval
$[1,2]$. As the pigeon hole bound has the air of suboptimality, we seek to replace it
with something tighter for this specific code.

\subsection{An example sequence of small determinants in Badr--Belfiore code}

In this section we study a sequence of determinants appearing in the BB-code
that converge towards zero. The example utilizes the fact that within the ring $\OO_2$
there are arbitrary small numbers. For example, because $|2-\sqrt5|\approx 0.2369<1/4$
its powers $(2-\sqrt5)^n=a_n-b_n\sqrt5$ can be made as small as required.

Let us consider the simple case $x_2=1$ and $x_1=a+b\sqrt5 i$,
where $a,b\in \Z$. Then
$x = x_1, \sigma(x) = a - b\sqrt5 i,$
so
$
\det(X)= x - i\sigma(x) = (a - b\sqrt5 )(1-i).
$

In order to make this as specific as possible
let us study the sequence of such matrices $X_n$  with $a=a_n$,
$b=b_n$, where for all $n>0$ the integers $a_n$ and $b_n$ are determined by the equation
$a_n-b_n\sqrt5 =(2-\sqrt5)^n.$ We remark that this is by no means the only sequence we could consider to achieve our goal. We can form other such sequences by multiplying this with constants and also use other small algebraic integers: any $(a,b)\in \Z^2$ pair such that $(a-b\sqrt5)$ is small will yield small determinants by this construction.

The number $\alpha=2+\sqrt5=\tau^3$ is a unit in the ring $\Z[\tau]$. Its norm is
$N^{F_2}_{\Q}(\alpha)=\alpha \sigma(\alpha)=(2-\sqrt5)(2+\sqrt5)=-1,$ and hence
$\sigma(\alpha)=2-\sqrt5 =-1/\alpha.$ This norm
equation gives us the identity
 $
a_n^2-5b_n^2=(-1)^n
$
 that is valid for all integers $n>0$. At this time we infer from this formula
that $|b_n|< |a_n|$ for all $n>0$.

We also have use for the trace function
$tr^{F_2}_{\Q}:F_2\rightarrow\Q,x\mapsto x+\sigma(x).$ For example, as
$\sigma(a_n-b_n\sqrt5)=a_n+b_n\sqrt5$ we get the formula
 $
2a_n=tr^{F_2}_{\Q}(\alpha^n)=\alpha^n+(-1/\alpha)^n.
$
 In this formula the second term always has absolute value $<1$, so the
first term dominates for large values of $n$, and
we get the asymptotic formula
 $
a_n\approx \frac{1}{2}(2+\sqrt5)^n.
$
 We shall also need an explicit expression of $b_n$ in terms of $\alpha$,
and the following formula is immediate from the definitions
 $
2\sqrt5 b_n=\alpha^n-(-1/\alpha)^n.
$

Now if we set in the BB-code $x_2=1$ and $x_1=z_n=a_n+i\sqrt5 b_n$, then
the logarithm of the resulting determinant looks like
$
\log|\det(X)|$ $=\log |(1+i)(2-\sqrt5)^n|$
$ = \log \sqrt2 + n \log|2-\sqrt5|$
$=\frac{\log 2}{2} -n \log\alpha$.

At the same time the range parameter $N$ grows as $\log N= n\log\alpha-\log2$. Therefore with this
example sequence we get the limit
 $
\lim_{n\rightarrow\infty}\frac{\log|\det(X)|}{\log{N}}=-1.
$ 

Thus this example sequence of matrices simply makes the single antenna pigeon hole bound explicit
for the BB-code.
The obvious route to a better upper bound for the decay function $D(N)$ of the BB-code is to
use this sequence of determinants, but to split the energy more evenly between the two users. After all, here (as in
our proof of the pigeon hole bound) one user was stuck with a low rate signal, while the other users data rate was
unbounded. To do this we want to write the numbers $z_n=a_n+i\sqrt5 b_n$ in the form
$z_n=x_1\sigma(x_2)$, where $x_1$ and $x_2$ would both use if not equal then at least comparable amounts of transmission power.
While we cannot do this for all the numbers $z_n$, a useful factorization exists, when $5|n$.
This is the topic of the following subsection.

\subsection{Certain factorizations in $\OO_E$}

Let $\zeta=e^{2\pi/5}$ be a fifth root of unity. Our field of interest $E$
is a subfield of the twentieth cyclotomic field $L=\Q(i,\zeta)$, and $[L:E]=2$.
This follows from the fact that $-\zeta-\zeta^{-1}=-2\cos (2\pi/5)<0$
is a zero of the polynomial $x^2-x-1=(x-\tau)(x-1+\tau)$, and hence
$
\tau=\zeta+1+\zeta^{-1}.
$

The degree $[L:\Q]=8$ follows from the fact that the minimal
polynomial of any primitive twentieth root of unity, such as $i\zeta$, is
 $
\phi_{20}(x)=x^8-x^6+x^4-x^2+1.
$ 
This is, perhaps, easiest to see starting with the factorization
 $
p(x):=x^{10}+1=(x^2+1)\phi_{20}(x).
$ 

There is an automorphism $\nu$ of $L$ that is determined by $i\mapsto i, \zeta\mapsto \zeta^{-1}$.
We immediately see that $\nu$ is of order two, and that $E$ is the fixed field of $\nu$.
So if $w$ is any root of unity of order 20, then the polynomial $(x-w)(x-\nu(w))$ has
coefficients in the field $E$. Using this we arrive at the following factorization of
$\phi_{20}(x)$ into irreducible factors in the ring $E[x]$:
 $
\phi_{20}(x)=p_1(x)p_2(x)p_3(x)p_4(x),
$ 
where
\begin{eqnarray*}
p_1(x)=&(x-i\zeta)(x-i\zeta^{-1})=x^2+i(1-\tau)x-1,\\
p_2(x)=&(x+i\zeta)(x+i\zeta^{-1})=x^2-i(1-\tau)x-1,\\
p_3(x)=&(x-i\zeta^2)(x-i\zeta^{-2})=x^2+i\tau x-1,\\
p_4(x)=&(x+ i\zeta^2)(x+ i\zeta^{-2})=x^2-i\tau x-1.
\end{eqnarray*}
The task at hand is to factorize the number $z_n=a_n+i\sqrt5 b_n$. The symmetries of
these numbers become more apparent, if we take a detour via $\Q$, so we start by considering
the $F_3\rightarrow\Q$ norm
 $
z_n\rho(z_n)$ $=a_n^2+5b_n^2$ $=a_{2n}$ $=\frac12 \left[\alpha^{2n}+\alpha^{-2n}\right].
$

As before, here $\alpha=2+\sqrt5$, so $\mu(\alpha)=\sigma(\alpha)=2-\sqrt5=-1/\alpha$.
The number $u=\alpha^{2n}$ will appear frequently in our calculations. We start our work
on $z_{5n}$ with
\begin{eqnarray*}
\frac{a_{10n}}{a_{2n}}=&\frac{\alpha^{10n}+\alpha^{-10n}}{\alpha^{2n}+\alpha^{-2n}}
=\frac{u^5+u^{-5}}{u+u^{-1}}=\frac{u^{-4}(u^{10}+1)}{u^2+1}\\
=&u^{-4}\phi_{20}(u)=m_1(n)m_2(n)m_3(n)m_4(n),
\end{eqnarray*}
where for $j=1,2,3,4$ we denote
 $
m_j(n)=u^{-1}p_j(u)=\alpha^{-2n}p_j(\alpha^{2n})\in \OO_E.
$

As $\mu(u)=1/u$, we have for example
$$
\mu(m_1(n))=\mu(u+i(1-\tau)-u^{-1})=u^{-1}-i\tau-u=-m_3(n).
$$
Similarly $\mu(m_2(n))=-m_4(n)$, and as $\mu^2=1$ in the Galois group
we get that $m_1(n)m_3(n)$ and $m_2(n)m_4(n)$ are invariant under $\mu$,
and hence are integers in the field $F_3$. Thus we may expect that
one of these pairs is a factor of $z_{5n}$.

We need one more pair of polynomial factorizations, this time in the ring $\OO_1=\Z[i];$
\begin{eqnarray}
x^5\pm i&=(x\pm i)(x^4\mp ix^3-x^2\pm ix+1).
\end{eqnarray}
These arise similarly from factoring $x^{20}-1$, or rather its factors $x^5+i$ and $x^5-i$ respectively, in $F_1[x]$.
They are needed  in the following lemma
that is the main result of this subsection.

\begin{lemma}\label{z5nfactorization}
The number $z_{5n}$ is always divisible by $z_n$, and it
can be factored in the ring $\OO_E$ as
 $
z_{5n}=z_n m_2(n) m_4(n),
$
 when $n$ is odd, and as
 $
z_{5n}=z_n m_1(n) m_3(n),
$ 
when $n$ is even.
\end{lemma}

\begin{proof}
Both of these identities follow from ther earlier expressions for $a_n$ and $b_n$ in
terms of powers of $\alpha$. These
may be compressed into formula
 $
z_n=\frac12 (1+i)(\alpha^n-i(-1/\alpha)^n).
$ 
Using our earlier abbreviation $u=\alpha^{2n}$ we see that
\begin{eqnarray*}
m_2(n)m_4(n)&=u^2-iu-1+iu^{-1}+u^{-2},\\
m_1(n)m_3(n)&=u^2+iu-1-iu^{-1}+u^{-2}.
\end{eqnarray*}
Let us consider the case $n$ odd. In this case we can write
$z_n=\alpha^{-n}(1+i)(u+i)/2.$ We also see that
$m_2(n)m_4(n)=\alpha^{-4n}(u^4-iu^3-u^2+iu+1)$. Therefore this case of the claim
follows from the first of the above polynomial factorizations by substituting
$x=u$. The even case follows similarly from the second polynomial factorization.
\end{proof}

\subsection{Sharper upper bounds to the decay function of the Badr--Belfiore code and numerical data}

Let us take a closer look at the factorization in Lemma \ref{z5nfactorization}.
We want to say something about the sizes of the coordinates of these algebraic integers with
respect to the integral basis $\{1,i,\tau,i\tau\}$. From all the previous identities it immediately
follows that the coordinates of the factors $m_j(n),j=1,2,3,4$, have absolute values bounded from above
by a constant multiple of $\alpha^{2n}$. Therefore the coordinates of $x_1=z_n m_j(n)$  ($j=1$ or $j=2$)
can be approximated by a constant multiple of $\alpha^{3n}$ and the coordinates of $x_2=\sigma(m_{j+2}(n))$
by a constant multiple of $\alpha^{2n}$. Recall that these choices yield a determinant
of absolute value $\sqrt2 \alpha^{-5n}$.

As any size parameter $N$ can be approximated up to a constant ($<\alpha^5$) multiplier with a power of $\alpha^5$
we have the following result.

\begin{corollary}
There exists such a constant $K>0$ that for all
$N$ the decay of the BB-code has an upper bound
$$
D(N^{3/5},N^{2/5})\le\frac{K}{N}.
$$
In particular, the decay exponent $\delta$ then has the estimates
$$
5/3\le\delta(\hbox{BB-code})\le 2.
$$
\end{corollary}

One way of getting better upper bounds for the decay exponent is to apply Lemma \ref{z5nfactorization}
repeatedly. After all, we get an even better balance between the factors $x_1$ and $x_2$, when $n$
is a multiple of 25, because in the factorization $z_{25n}=z_{5n} m_j(5n)m_{j+2}(5n)$ we can
factor $z_{5n}$ further.

Observe that when doing this, we effectively restrict our scale
to the sizes $a_1, a_5, a_{25}, a_{125},\ldots$. Thus we lose the ability to estimate (up to a constant multiplier)
an arbitrary scale parameter $N$ by a member of this sequence. Therefore the following result is stated in terms of
limes superior.

\begin{corollary}
For the BB-code we get the result
$$
\limsup_{N\rightarrow\infty}-\frac{\log D(N,N)}{\log N} = 2.
$$
\end{corollary}

We conclude this section by a table of numerical results based on the above factorization.
Two things are obvious. The multiples of 25 stand out. Note also that the coordinates of these
factors are quite large (but the determinant is then correspondingly very small), and surely
beyond the range of all ongoing simulations.

\begin{table}[h!]\caption{Some small determinants in BB-code and estimates of $\delta$}\vspace{-0.4cm}
\label{p1mod20factors}
\begin{center}
\renewcommand{\arraystretch}{1.22}
 \begin{tabular}{|c|c|c|} \hline
 \ $n$\ &\ $m=$ max size of $x_i$ a factor of $z_n$\ &\ $\delta=-\log \det(X)/\log m$  \\ \hline
 5 & 38 & 1.889\\
10 & 2880 & 1.769\\
15 & 219640 & 1.732\\
20 & 16692480 & 1.715\\
25 & 66563198 & 1.984\\
 \hline
\end{tabular}
\end{center}
\end{table}

\section{DMT Performance of the Badr-Belfiore Code} \label{sec:4}

Recall that in Section II, the rows of the BB code are formed by the lattices associated with each user with coordinates $a_j, b_j, c_j, d_j$, $j=1,2$ lying within the range of $[-N,N]$. Thus, following from \eqref{eq:N_j}, assuming the users are to achieve multiplexing gain $r_1=r_2=r$, the corresponding value for $N$ is 
\[
N \ = \ \text{SNR}^{\frac{r}{2}}
\]
since $n_t=1$ is used in the BB code. Furthermore, as the elements $\tau$ and $\gamma$ are fixed and do not vary with SNR, it is straightforward to see that the overall BB-code matrix $X$ in \eqref{eq:Xbb} has average power 
${\mathbb E} \left\| X \right\|^2 \ \dot\leq \ N^2 \ = \ \text{SNR}^{r}$. 

In \cite{Cor} Coronel et al. had provided some initial DMT analysis of the BB code. They showed that the BB code will be MAC-DMT optimal if the following inequality is satisfied 
\begin{equation}
2 r + \delta \ \leq \ r_{\cal S} \left( d_{{\cal S}^*} \left( r \left( {\cal S}^* \right) \right) \right) \label{eq:cond}
\end{equation}
where $r_{\cal S} \left( d_{{\cal S}^*} \left( r \left( {\cal S}^* \right) \right) \right)$ is the maximum of the sum of multiplexing gains of users in set ${\cal S}$ such that the dominant diversity gain $d^*(r)=d_{{\cal S}^*} \left( r \left( {\cal S}^* \right) \right) = \max\left\{ d_{1,2}^*(r), d_{2,2}^*(2r) \right\}$ can be achieved. ${\cal S}^*$ is the set of the users that is dominant in the DMT error performance. Specifically, ${\cal S}^*=\{1\}$ for $r \in [0,\frac{2}{3}]$ and is called single-user performance region in \cite{TseVisZhe}. For 
$r \in \left[\frac{2}{3}, 1 \right]$  we have ${\cal S}^*=\{1,2\}$ and this is termed the antenna-pooling region. $d_{p,q}^*(x)$ is the point-to-point DMT with $p$ transmit and $q$ receive antennas given multiplexing gain $x$ given in \cite{ZT}. Note that $d_{1,2}^*(x)=2-2x$ for $x \in [0,1]$ and $d_{2,2}^*(x)=4-3x$ for $x\in[0,1]$ and $d_{2,2}^*(x)=2-x$ for $x \in [1,2]$. To achieve diversity gain $d^*(r)=2-2r$, it is easy to show for ${\cal S}=\{1,2\}$ we have 
\[
r_{\cal S} \left( d_{{\cal S}^*} \left( r \left( {\cal S}^* \right) \right) \right) \ = \ \left\{
\begin{array}{ll}
\frac{2+2r}{3}, & r \in \left[0,\frac{1}{2}\right]\\
2r, & r \in \left[\frac{1}{2}, 1 \right].
\end{array} \right. 
\] 
The other parameter $\delta$ shown in \eqref{eq:cond} is defined as 
\[
\delta \ := \ - \limsup_{\text{SNR} \to \infty}\ \log_{\text{SNR}} \min_{X \neq X'} \left| \det \left( X-X' \right) \right|^2 
\]
where $X$ and $X'$ are distinct overall matrix of the BB code. In terms of the notion $D(N,N)$ we have as $\text{SNR} \to \infty$
\[
\delta \ = \ - \log_{\text{SNR}} \left| D(N,N) \right|^2\ = \  \log_{\text{SNR}} N^4 \ = \ 2r
\]
where the second equality follows from Corollary 3.3 and where we have set $N = \text{SNR}^{\frac{r}{2}}$ such that both users achieve multiplexing gain $r$. Putting all of the above together into \eqref{eq:cond} shows that the BB code is MAC-DMT optimal when the multiplexing gain falls in the interval of $\left[0,\frac{1}{5}\right]$, but fails to achieve the condition \eqref{eq:cond} by Coronel et al. for $r \geq \frac{1}{5}$. We summarize the above in the following result. 
\begin{thm}
The BB-code is MAC-DMT optimal when the multiplexing gain $r \leq \frac{1}{5}$. 
\end{thm}


\end{document}